\documentclass[a4paper]{jpconf}
\usepackage{amsmath,amsfonts,amssymb}
\usepackage{graphicx,bm,color,subfigure,hyperref,xspace,multirow}
\usepackage{hyperref}
\usepackage[normalem]{ulem}

\newcommand{\ie}{{\it i.e.}\xspace}
\newcommand{\abs}[1]{\ensuremath{|#1|}}
\DeclareMathOperator{\im}{Im}
\DeclareMathOperator{\re}{Re}
\newcommand{\SigmaD}{\ensuremath{\Sigma_c^+ \bar{D}^0}\xspace}
\newcommand{\Pc}{\ensuremath{P_c(4312)^+}\xspace}
\newcommand{\Pca}{\ensuremath{P_c(4440)^+}\xspace}
\newcommand{\Pcb}{\ensuremath{P_c(4457)^+}\xspace}
\newcommand{\jpsi}{\ensuremath{J/\psi}\xspace}
\newcommand{\jpsip}{\ensuremath{J/\psi\,p}\xspace}
\newcommand{\mevnospace}{\ensuremath{{\mathrm{\,Me\kern -0.1em V}}}}
\newcommand{\gevnospace}{\ensuremath{{\mathrm{\,Ge\kern -0.1em V}}}}
\newcommand{\tevnospace}{\ensuremath{{\mathrm{\,Te\kern -0.1em V}}}}
\newcommand{\mev}{\mevnospace\xspace}
\newcommand{\gev}{\gevnospace\xspace}

\begin{document}
\title{On the possible virtual state nature of the LHCb \Pc signal}

\author{J.~A.~Silva-Castro}

\address{Instituto de Ciencias Nucleares, 
Universidad Nacional Aut\'onoma de M\'exico, Ciudad de M\'exico 04510, M\'exico}

\ead{jorge.silva@correo.nucleares.unam.mx}

\begin{abstract}
In this contribution we study the nature of the new 
\Pc signal reported by LHCb collaboration in the \jpsip spectrum. 
We use $S$-matrix principles to perform a minimum-bias 
analysis of the data,  
focusing on the analytic properties that can be related to the microscopic 
origin of the \Pc peak.
Using the scattering length approximation
we find evidence for interpretation of the signal
as a virtual state generated by the 
attractive effect of the \SigmaD channel opening.
\end{abstract}



\section{Introduction}
   Searching for and understanding of exotic hadron states, say, 
those that go beyond the minimal quark model
combination of three quarks for baryons and
a quark-antiquark pair for mesons, has become a
relevant topic  in physics over the recent years, because of the insight that these 
states can provide on the strong interaction
in the hadronic
energy regime~\cite{Esposito:2017Multiquark,Olsen:2018NonSMenBary,Guo:2017jvc}.
These states are allowed within Quantum Chromodynamics (QCD), as the fundamental theory of the strong interaction only requires that hadrons are color singlets without limiting the 
number of quarks. In this way, baryons that go beyond the three quark picture are not precluded by QCD.

In the baryon sector, the only exotic candidates 
so far are
the hidden-charm pentaquarks
recently found by the LHCb collaboration in the
$\Lambda_b^0 \to \jpsip \,K^-$ decay~\cite{Aaij:2015Pc,Aaij:2016Pc,Aaij:2019vzc},
labeled as \Pc, $P_c(4380)^+$, \Pca, and \Pcb.
In this work we focus our attention in the \Pc 
candidate~\cite{Aaij:2019vzc}, which is approximately $5\mev$ below the \SigmaD threshold.
There are several interpretations 
of this signal, such as a \SigmaD molecule 
\cite{Wu:2010jy,Wu:2010vk,Wang:2011rga,Yang:2011wz,Xiao:2013yca,Yamaguchi:2017zmn,Liu:2019Pc,Burns:2019Pc,Du:2019pij},
a compact pentaquark 
\cite{Zhu:2015bba,Ali:2019npk,Holma:2019lxe,Maiani:2015vwa}, and a virtual state as 
we suggest~\cite{Cefera:2019Pc4312}.\footnote{A virtual state is produced for example by an attractive 
interaction which is not strong enough to bind a state, as in the
neutron-neutron scattering~\cite{Frazer:1964zz,Hammer:2014rba}.}
To reach the conclusion that the virtual state interpretation was a plausible one, we used
a data-driven approach based on $S$-matrix theory
and minimally biased methods as shown in the next section. The detailed description of the method and the full
results can be found in \cite{Cefera:2019Pc4312}.


\section{Analysis of the \Pc region}

    A thoroughgoing analysis of the \Pc signal, for example, to determine its 
quantum numbers would require a full six-dimensional amplitude analysis fitting both 
the energy and angular dependencies. 
Nonetheless, the manifestation of the this signal in the experimental data allows a one 
dimensional analysis because it stands out as a narrow peak ($\sim 10\mev$) in what 
otherwise appears to be a smooth background.
The closeness of the \SigmaD channel opening
indicates that it could be the driving effect 
behind the appearance of the peak.
If it is so, the \Pc would likely be either
a molecule or a virtual state.

Thereby, we consider a coupled channel amplitude between 
the \jpsip and \SigmaD channels and restrict the analysis to the 4250-4380 \mev region
where the \Pc signal is found. 
We assume that the \Pc signal has a well defined 
spin \ie, it appears in a single partial wave $F(s)$, furthermore, the background 
from all other partial waves $B(s)$ is added incoherently and parametrized as a linear
polynomial $B(s)= b_0+b_1s$, which also encodes the contribution from further singularities. 
\footnote{
More formally, an analytical function can be expanded
into a Laurent series around a singularity in $a$ as
$f(z)=\frac{b}{z-a}+c_0+c_1z+c_2z_2+\cdots$, where $b$
is the residue of the pole and $c_i$  coefficients, 
far away from the singularity only the polynomial part contributes to the signal.
With a linear polynomial in our analysis we have $\chi^2/\textrm{dof}\sim 0.8$ 
with no improvements to higher orders.}

Hence, the events distribution is given by:
\begin{equation}
\frac{dN}{d\sqrt{s}}= \rho(s) \left[ \abs{F(s)}^2 + B(s) \right]
= \rho(s) \left[ \abs{P_1(s)T_{11}(s)}^2 + b_0 + b_1s \right]
,\label{eq:events}
\end{equation} 
where $\rho(s)$ is the phase space factor for the decay $\Lambda_b^0 \to \jpsip\, K^-$, 
given by $\rho(s) =m_{\Lambda_b} p\, q$  with 
 $p = \lambda^{1/2}(s,m^2_{\Lambda_b},m^2_K)/2m_{\Lambda_b}$ and
 $q = \lambda^{1/2}(s,m^2_p,m^2_\psi)/2\sqrt{s}$,
where $\lambda(x,y,z) = x^2 + y^2 + z^2 - 2xy - 2xz - 2yz$ is the K\"all\'en function.
This expression assumes that the signal is on the $S$-wave, and remains valid even 
if the signal is on another $\ell$ wave. This would imply adding a term $q^\ell$ in front of 
$F(s)$, which in practice remains constant due to $q$ does not change on the energy 
range considered.

   The function $F(s)$ is a product of a function $P_1(s)$ which provides
the production of $\jpsip\,K^-$ and also takes into account the effect of other signals 
projected onto the same partial wave of the \Pc, and the $T_{11}(s)$ amplitude, which
describes the $\jpsip\to\jpsip$ scattering, where the Pc is.

Near the \SigmaD threshold the two coupled channel
$T$ matrix can be written as \cite{Frazer:1964zz}:
\begin{subequations}
	\begin{align}
	T_{11}(s)&=\frac{M_{22}-ik_2}{(M_{11}-ik_1)(M_{22}-ik_2)-M_{12}^2},\label{eq:11}\\
	T_{12}(s)&=\frac{-M_{12}}{(M_{11}-ik_1)(M_{22}-ik_2)-M_{12}^2},\\
	T_{22}(s)&=\frac{M_{11}-ik_1}{(M_{11}-ik_1)(M_{22}-ik_2)-M_{12}^2},
	\end{align}
\end{subequations}
where  $k_i = \sqrt{s - s_i}$, and  $s_1 = (m_{\psi} + m_{p})^2$,
$s_2 = (m_{\Sigma^+_c} + m_{\bar{D}^0})^2$ are the thresholds of the two channels.
In this way, $T_{11}(s)$ represents the reaction $\jpsi p\to\jpsip$, 
$T_{22}(s)$ the reaction
$\SigmaD\to\SigmaD$ and the off diagonal terms
the chanels $T_{12}(s)$ $\jpsip\to\SigmaD$ and $T_{21}$ $\SigmaD\to\jpsip$.

Due to the unitarity condition, the elements of the real symmetric  $2\times 2$ 
matrix $M(s)$ are singularity free and can be Taylor expanded. In this work we focus 
in the scattering length approximation which results from keeping only the first 
term. The first-order effective range expansion, say $M_{ij}(s) = m_{ij}  - c_{ij} s$,
is discussed in \cite{Cefera:2019Pc4312}. The function $P_1(s)$ is analytic in the data 
region, and, given the small mass range considered, it can be parametrized with a 
first order polynomial $P_1(s)=p_0+p_1s$. The 
$M_{12}$ parameter is linked to the
channel coupling. 
We  stress that, since the $\jpsip$ threshold is far away from the region of interest, 
this channel can effectively absorb all the other channels with distant thresholds.
In principle we should add the 
off-diagonal term $P_2(s)T_{12}(s)$, but, this would have
no impact in the analytic properties and poles
of the amplitude. In our case
we ignore such a term to reduce the number of free parameters.

%
\begin{figure}
	\centering
	\subfigure[\ ]{
		\includegraphics[width=.48\textwidth]{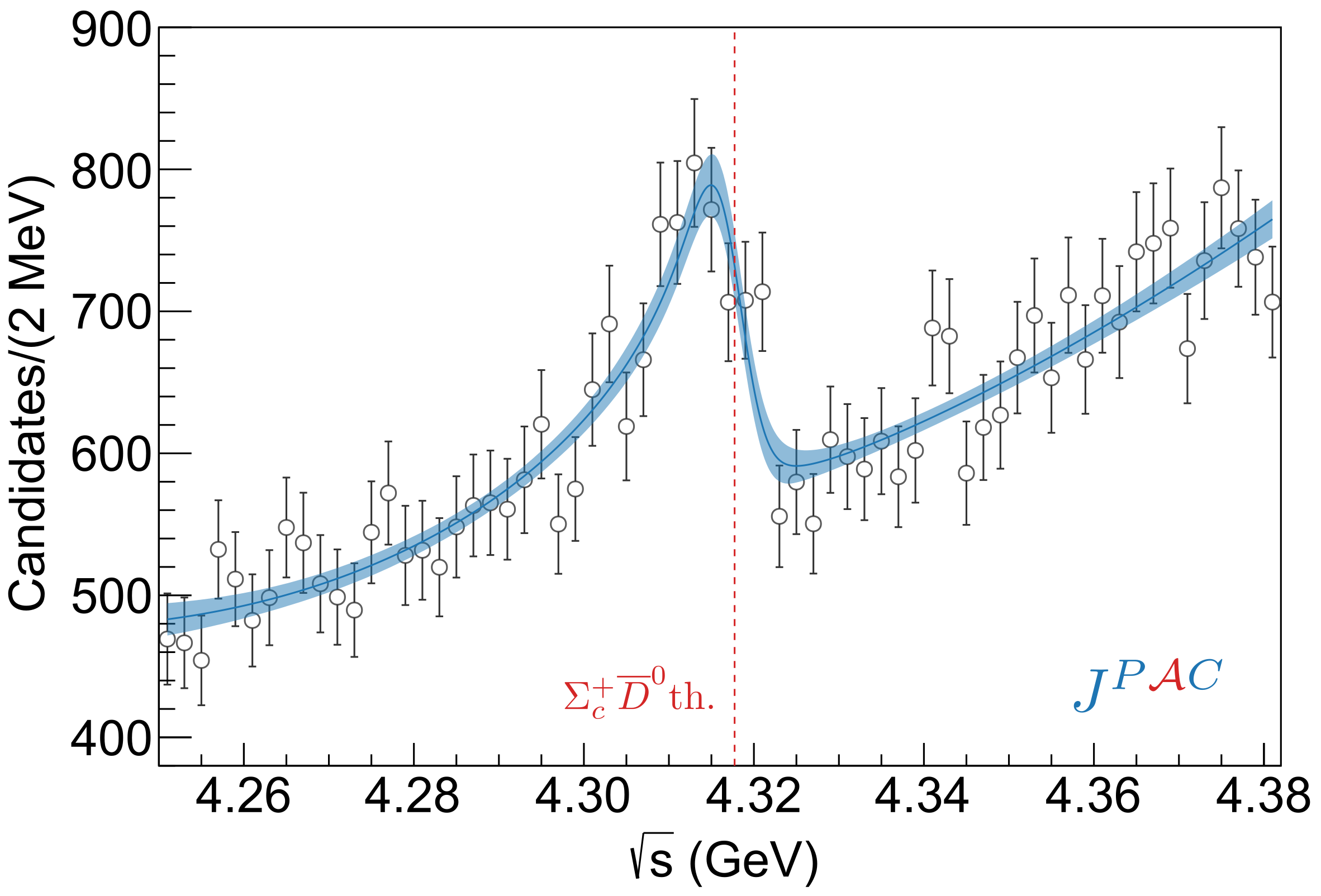}
		\label{fig:data}}
	\subfigure[\ ]{
		\includegraphics[width=.48\textwidth]{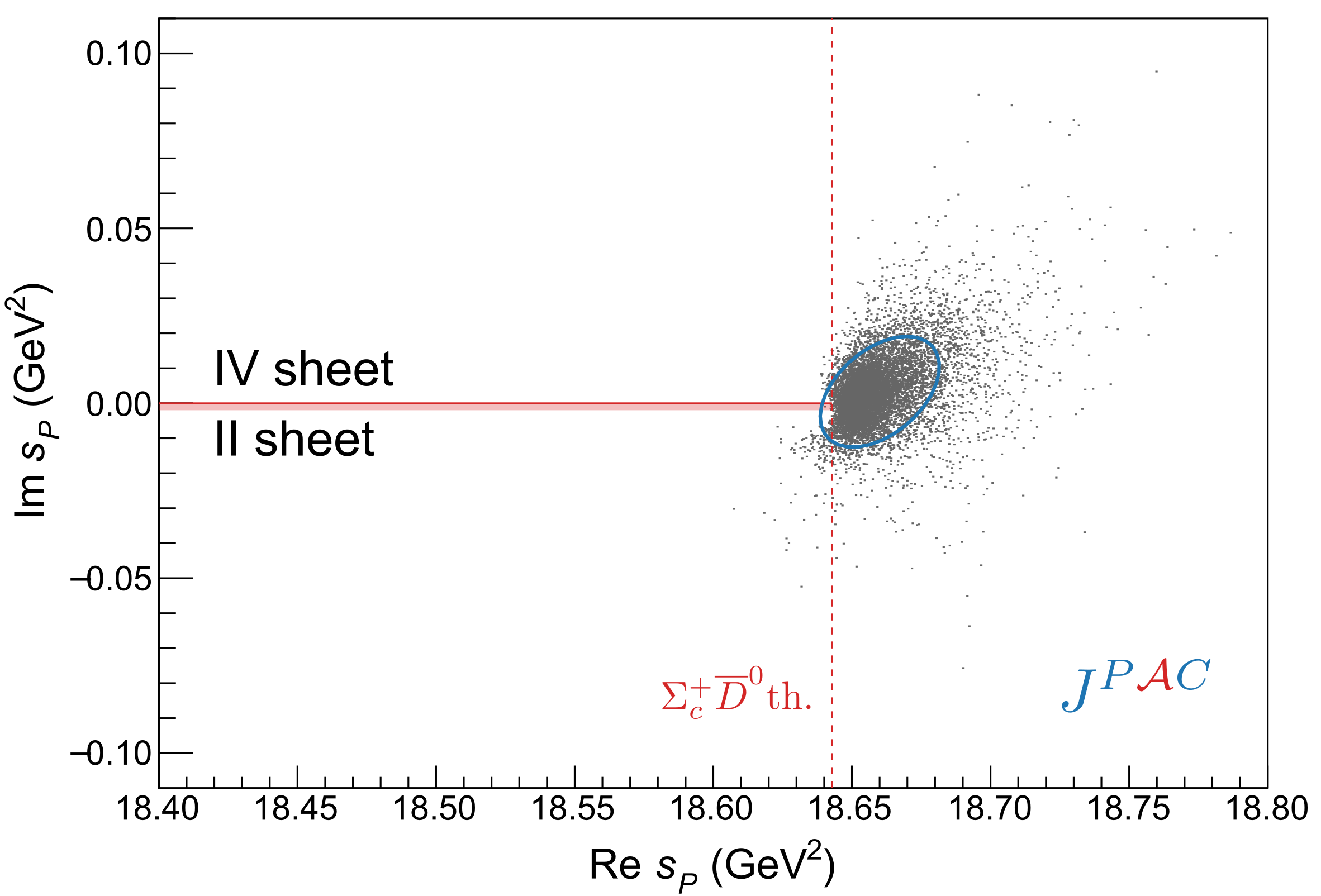}
		\label{fig:pole}}
	\caption{ (a) Fit to the $\cos\theta_{P_c}$-weighted \jpsip mass 
	    distribution from 
        LHCb~\cite{Aaij:2019vzc} in the scattering length approximation, 
        (equations \eqref{eq:events} and \eqref{eq:11}). 
        The solid line and blue
        band show the result of the fit and the $1\sigma$ 
        confidence level provided by the bootstrap analysis respectively.
        (b)
        Poles obtained from the $10^4$ bootstrap fits
        in the scattering length approximation. The physical region is highlighted with a 
        pink band. 
        For each bootstrap fit only one pole appears in this region 
        and the blue ellipse accounts the 68\% of the cluster concentrating 
        above threshold.}\label{fig:fits}
\end{figure}
%


\section{Results and discussion}

    Due to the square roots in the denominator, the amplitude in Eq.~\eqref{eq:11}
has branch cuts opening at the two thresholds. It turns out that there are four
Riemann sheets, and in the scattering length approximation any pole can only appear on 
either the II or the IV sheets~\cite{Frazer:1964zz}. 
This parameterization allows for the 
description of bound molecules and unbound virtual states. When we turn off 
the coupling between the two channels, \ie $M_{12}\to 0$, a molecular interpretation 
occurs if the pole moves to the real axis of the physical sheet below the heavier 
threshold, and a virtual state occurs if the pole moves onto the real axis of the 
unphysical sheet (see also Fig. 2 of Ref. \cite{Pilloni:2016obd} and the corresponding
description).
We note that the pole movement is model dependent, as
the parameters are not independent but related by the underlying QCD dynamics.
This is a problem common to every model or effective hadron theory.
    
    We fit the $\cos\theta_{P_c}$-weighted spectrum $dN/d\sqrt{s}$ measured in 
\cite{Aaij:2019vzc}, with $\sqrt{s}$ being the \jpsip invariant mass, using MINUIT
\cite{minuit}, considering the experimental resolution reported in
\cite{Aaij:2019vzc}.
This spectrum is obtained  by applying $\cos\theta_{P_c^+}$-dependent weights 
to each candidate to enhance the $P_c^+$ signal, where $\theta_{P_c}$ is the angle between 
the $K^-$ and $\jpsi$ in the $P_c^+$ rest frame (the $P_c^+$ helicity angle 
\cite{Aaij:2019vzc,Aaij:2015tga}).

To estimate the sensitivity of the pole positions 
to the uncertainties in the data, we use the bootstrap technique~\cite{recipes,EfroTibs93}; 
\ie we generate $10^4$ pseudodata sets and fit each one of them. 
The statistical fluctuations in data reflect into the the uncertainty band 
plotted in Fig.~\ref{fig:data}. Moreover, for each of these fits, we determine 
the pole positions, as shown in Fig.~\ref{fig:pole}.

    In this analysis it is possible to identify a cluster of virtual state poles 
across the II and IV sheet above the \SigmaD threshold (see also the discussion in Ref.~\cite{Rodas:2018owy}). 
If we use the customary definition of mass and width,
$M_P = \re \sqrt{s_p}$, $\Gamma_P = -2\im \sqrt{s_p}$  the main cluster has 
$M_P = 4319.7 \pm 1.6\mev$, $\Gamma_P = -0.8 \pm 2.4\mev$, where positive or 
negative values of the width correspond to II or IV sheet poles, respectively. 
To establish the nature of this singularity,  we track down the movement of the poles as 
the coupling between the two channels is reduced. By taking 
$m_{12} \to 0$, we can see how the cluster moves over to the upper side of the IV 
sheet and ends up on the real axis below the \SigmaD threshold~\cite{pcjpaclink}. 
The fraction of poles that reach the real axis from the lower side of the II 
sheet is $0.7\%$ only, and thus not significant.
This result reinforces the interpretation of the pole as an unbound virtual state, 
meaning that the binding between the $\Sigma_c^+$ baryon and the $\bar D^0$ meson is enough to generate a signal,
but insufficient to form a bound molecule. 

As a cross-check we also analyze the unweighted \jpsip spectrum in the same region, 
both with and without the $m_{Kp} > 1.9\gev$ cut. 
All the results are consistent.

\section{Conclusions}
    We have studied the \Pc signal reported by LHCb in the \jpsip spectrum by 
considering a reaction amplitude which satisfies the general principles of 
$S$-matrix theory, restricting the analysis to the scattering length approximation.
    The analytic properties of the amplitudes can be related to the microscopic 
origin of the signal. We fitted the LHCb mass spectrum in the $4312\mev$ mass region 
including the experimental resolution. 
The statistical uncertainties in the data were propagated to the extracted poles 
using the bootstrap technique. 
We do not find support for a bound molecule, we conclude that the most likely interpretation 
of the \Pc peak is a virtual (ubound) state.


\ack
This  work  is  part  of  the  efforts  of  the  Joint  Physics  Analysis  Center  
(JPAC) collaboration.
The author thanks his colleagues and co-authors of the original publication:
C\'esar Fern\'andez-Ram\'irez, Alessandro Pilloni, Miguel Albaladejo, Andrew Jackura, Vincent Mathieu, Mikhail Mikhasenko, 
and Adam Szczepaniak.
This work was supported by
PAPIIT-DGAPA (UNAM, Mexico) Grant No.~IA101819, 
and CONACYT (Mexico) Grants No.~734789 and~No.~A1-S-21389.
The author also thanks the organizers for their invitation to the conference and their warm hospitality at Cocoyoc.


\section*{References}
\bibliographystyle{iopart-num}
\bibliography{ref.bib}

\end{document}